\documentclass[12pt]{article}
\usepackage{graphicx,psfrag,epsfig}
\usepackage{amssymb,amsmath,amscd,amsthm}
\usepackage{graphicx,psfrag,epsfig}

\usepackage{graphicx}
\usepackage[active]{srcltx}

\newtheorem{theorem}{Theorem}[section]
\newtheorem{corollary}[theorem]{Corollary}

\newtheorem{proposition}[theorem]{Proposition}

\date{}

\begin{document}

\date{}
\title{On the negative spectrum of the hierarchical Schr\"{o}dinger operator.}
\author{
 S.
Molchanov\footnote{Dept of Mathematics, University of North
Carolina, Charlotte, NC 28223, smolchan@uncc.edu; corresponding author},
B. Vainberg\footnote{Dept of Mathematics, University of North
Carolina, Charlotte, NC 28223, brvainbe@uncc.edu}
}
 \maketitle
\begin{abstract}
This paper is devoted to the spectral theory of the Schr\"{o}dinger operator on the simplest fractal: Dyson's hierarchical lattice. An explicit description of the spectrum, eigenfunctions, resolvent and parabolic kernel are provided for the unperturbed operator, i.e., for the Dyson hierarchical Laplacian. Positive spectrum is studied for the perturbations of the hierarchical Laplacian. Since the spectral dimension of the operator under consideration can be an arbitrary positive number, the model allows a continuous phase transition from recurrent to transient underlying Markov process. This transition is also studied in the paper.
\end{abstract}
 

\section{Introduction}

The spectral theory of the fractals, which are similar to the infinite Sierpinski gasket (i.e. the spectral theory of the corresponding Laplacians) is well understood (see \cite{fuk}, \cite{mal}, \cite{kig}). It has several important features: the existence of a large number of eigenvalues of infinite multiplicity, pure point structure of the integrated density of states, compactly supported eigenfunctions. These features manifest themselves in the unusual asymptotics of the heat kernel, the specific structure of the corresponding $\zeta$-function, etc., see \cite{acc}.

The next natural step in the spectral theory is to study Schr\"{o}dinger type operators, i.e., fractal Laplacian perturbed by a potential. There are two possible directions for such a development: analysis of the random Anderson Hamiltonians (the potential is stationary in space) or the study of the classical problem on the negative spectrum when the potential vanishes at infinity. For the first direction, see \cite{bov}, \cite{md}, \cite{kri}. We will concentrate on the second problem in a particular case of the simplest fractal object: Dyson's hierarchical Laplacian perturbed by a decaying potential. Our goal is to prove the Cwikel-Lieb-Rozenblum (CLR) estimates for the number of negative eigenvalues and estimates for Lieb-Thirring (LT) sums. These estimates depend on the spectral dimension $s_{h}$ of the fractal (which can take an arbitrary positive value.) The most important part of the paper is the analysis of the spectral bifurcation near the critical dimension $s_{h}=2$.

The authors are very grateful to E. Akkermans, J. Avron and A. Teplyaev for useful discussions.

\section{Hierarchical lattice and Laplacian}

The concept of the hierarchical structure was proposed by F. Dyson \cite{Dy} in his theory of 1-D ferromagnetic phase transitions. There are several modifications of the hierarchical Laplacian (see \cite{md}). We will study the simplest one, which is characterized by an integer-valued parameter $\nu\geq 2$ and a probabilistic parameter $p \in( 0,1)$.

\textbf{Description of the model.}
Consider a countable set $X$ and a family of partitions $\Pi_0\subset\Pi_1\subset\Pi_2\subset...$ of $X$ (we write ${\Pi _r} \subset {\Pi _{r + 1}}$ to mean that every element of ${\Pi _r}$ is a subset of some element of ${\Pi _{r + 1}}$). The elements of $\Pi_0$ are the singleton subsets of $X$. They are denoted by $Q_i^{(0)}$ and called cubes of rank zero. Each element $Q_i^{(1)}$ of $\Pi_1$ (cube of rank one) is a union of $\nu$ different cubes of rank zero, i.e., $X=\cup Q_i^{(1)},~| Q_i^{(1)}|=\nu$ (see Figure 1). Each element $Q_i^{(2)}$ of $\Pi_2$ (cube of rank two) is a union of $\nu$ different cubes of rank one, i.e., $X=\cup Q_i^{(2)},~| Q_i^{(1)}|=\nu^2$, and so on. The parameter $\nu\geq 2$ is one of the two basic parameters of the model.

\begin{figure}[!ht]
\vspace{0.3 cm}
\centering
\includegraphics[width=.6\textwidth]{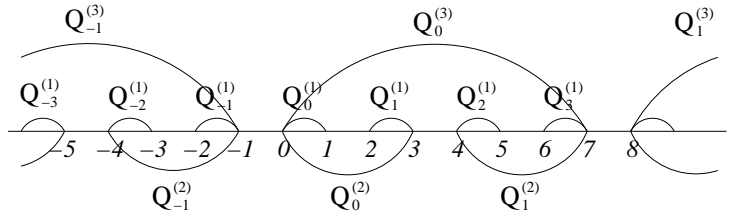}
\caption{An example of a hierarchical lattice with $X = \mathbb{Z}$ and $\nu=2.$}
\label{Picture1}
\end{figure}

Each point $x$ belongs to an increasing sequence of cubes of each rank $r\geq 0$ which we denote by $Q^{(r)}(x)$, i.e., $x = Q^{(0)}(x)\subset Q^{(1)}(x)\subset Q^{(2)}(x)\subset\cdots~$.

 The hierarchical distance $d_h(x,y)$ on $X$ is defined as follows:
\begin{equation}\label{dh}
d_h(x,y)=\min\{r:\exists Q^{(r)}_i\ni x,y\}.
\end{equation}
We assume the following connectivity condition holds: for each $x,y\in X$, the cubes $Q^{(n)}(x)$ contain $y$ when $n$ is large enough, i.e., $d_h(x,y)<\infty.$

Note that for arbitrary $z\in X, ~ d_{h}(x,y)\leq \max \left\{d_{h}( x,z) ,d_{h}(y,z) \right\}$, i.e., $d_{h}(\cdot ,\cdot) $ is a super-metric which implies that
\begin{equation*}
\rho(x,y) =\rho_{\beta }(x,y)=e^{\beta d_{h}\left( x,y\right) }-1,~\beta>0,
\end{equation*}
is also a metric. We will use it in the form
\begin{equation}\label{dist}
\rho(x,y) =\left( \frac{1}{\sqrt{p }}\right) ^{d_{h}\left(x,y\right) }-1,
\end{equation}
i.e., $\beta =\ln \frac{1}{\sqrt p }$. Here $p \in (0,1) $ is the second parameter of the "Laplacian" ${\Delta _h}$ (see formula (\ref{hlap}) below).

Now we denote by $l^{2}\left( X\right) $ the standard Hilbert space of square summable functions on the set $X$ and define a self-adjoint bounded operator (the hierarchical Laplacian) depending on the parameter $p \in \left( 0,1\right) :$%
\begin{equation}\label{hlap}
\Delta_{h}\psi \left( x\right) =\sum\limits_{r=1}^{\infty }a_{r}
\left[ \frac{\sum\limits_{x^{\prime }\in Q^{\left( \nu\right) }\left(
x\right) }\psi (x')}{\nu^{r}}-\psi \left( x\right) \right],\quad \text{where} \quad a_r=(1-p)p^{r-1}, \quad \sum_{r=1}^\infty a_r=1.
\end{equation}%

The random walk on $(X,d_h)$ related to the hierarchical Laplacian has a simple structure. It spends an exponentially distributed time $\tau$ (with parameter one) at each site $x$. At the moment $\tau +0$ it randomly selects the rank $k$ of a cube $Q^{(k)}(x),~k\geq 1,$ with $P\{k=r\}=a_r$ and jumps inside of $Q^{(k)}(x)$ with the new position $x'\in Q^{(k)}(x)$ being uniformly distributed.

It is clear that $\Delta_{h}=\Delta_{h}^*,~\Delta_{h}\leq 0,$ Sp$(\Delta_{h})\in[-1,0]$. The following decomposition will play an essential role. Denote by $I_K(x)$ the indicator function of a set $K\in X$, i.e., $I_K=1$ on $K$, $I_K=0$ outside of $K$. Then, for each $y\in X$,
\begin{equation}\label{ddec}
\delta_y(x)= \sum\limits_{k = 1}^\infty  {\left( {\frac{{{I_{{Q^{(k - 1)}}(y)}}(x)}}{{{\nu ^{k - 1}}}} - \frac{{{I_{{Q^{(k)}}(y)}}(x)}}{{{\nu ^k}}}} \right)}.
\end{equation}
The validity of (\ref{ddec}) is obvious. It is important that each term on the right is an eigenfunction of $\Delta_{h}$ and the $k^{\text{th}}$ term belongs to the eigenspace $L_k$ defined in the following proposition.

\begin{proposition}\label{prd1}
$(a)$ The spectrum of $\Delta_{h}$ consists of isolated eigenvalues $\lambda_k=-p^{k-1},~k=1,2,...$, each of infinite multiplicity, and their limiting point $\lambda=0$.

$(b)$ The corresponding eigenspaces $L_k \subset l^2(X)$ have the following structure: For $k=1$,
$$
L_1=\{\psi\in l^2(X):~\sum\nolimits_{x\in Q_i^{(1)}}{\psi (x)}=0 \quad {\text{for each}} \quad Q_i^{(1)}\in \Pi_1 \}.
$$
For $k>1$, the space $L_k$ consists of all $\psi\in l^2(X)$ which are constant on each cube $Q_{i}^{(k-1)}$, and have the property that $\sum_{x\in Q_i^{(k)}}\psi(x)=0$ for each $Q_i^{(k)}\in \Pi_k$.

$(c)$ The following decomposition holds:   $l^2(X)=\oplus_{r=1}^\infty L_r.$
\end{proposition}
Indeed, one can easily check that the space $L_k$, defined above, consists of eigenfunctions with the eigenvalue $\lambda_k=-p^{k-1}$, and for each $y\in X$, the $k^{\text{th}}$ term in (\ref{ddec}) belongs to $L_k$. Thus (\ref{ddec}) immediately implies (c) which justifies (a).

Let us note that each eigenspace $L_k$ has an orthogonal basis of compactly supported eigenfunctions. Such a basis in $L_1$ consists of functions which are zero outside of a fixed cube $Q_{i }^{(1) }$ and such that $\sum_{x\in Q_{i }^{(1) }}\psi(x)=0$. There are $\nu-1$ orthogonal functions with the latter property for each cube $Q_{i }^{(1) }$. The orthogonal complement of $L_1$ consists of the functions $\psi \in l^2(X)$ which are constant on each cube of rank one. The basis in $L_2$ is formed by functions supported by individual cubes of rank two such that $\psi(x)=c_i$ on sub-cubes $Q_{i}^{(1) }$ of rank one, and $\sum c_i=0.$ One needs to specify $c_i$ to guarantee the orthogonality of the elements of the basis. The basis in $L_k,~k>1,$ is formed by functions which are supported by individual cubes of rank $k$ and which are constant on sub-cubes of rank $k-1$ with the sum of those constants being zero.

Let's find the density of states for $\Delta _{h}$ and the spectral dimension $s_{h}.$ We fix $x_0\in X$ (the origin) and a positive integer $N$. Consider the spectral problem
\begin{equation*}
-\Delta _{h}\psi =\lambda \psi;~~\psi \equiv 0{\text{ on }}X\backslash Q^{(N)}( x_{0}).
\end{equation*}%
(now it is more convenient to work with $-\Delta _{h}$ instead of $
\Delta _{h})$. It is easy to see (compare to Proposition \ref{prd1}) that the problem has the
following eigenvalues:
\begin{eqnarray*}
\lambda _{0,N} &=&1~~\text{ with multiplicity }\nu^{N-1}( \nu-1) \\
\lambda _{1,N} &=&p~~\text{ with multiplicity }\nu^{N-2}( \nu-1) \\
.......\\
\lambda _{N-1,N} &=&p^{N-1}~~\text{ with multiplicity }( \nu-1) \\
\lambda _{N,N} &=&p^{N}~~\text{ with multiplicity }1.
\end{eqnarray*}

This implies the following relation for
$$
\mathcal N_N(\lambda)=\frac{1}{\nu^N}\#\{\lambda_{i,j}<\lambda\}.
$$

\begin{proposition}\label{pr2} As $N\rightarrow \infty $,

\begin{equation*}
\mathcal N_N(\lambda) \rightarrow N(\lambda)=\sum\limits_{k\geq 0:p^{k}<\lambda }\frac{1}{\nu^{k}}\left( 1-\frac{1}{\nu}\right)=\frac{1}{\nu^{k_{0}\left( \lambda \right) }},
\end{equation*}
where ${k_0}(\lambda ) = \min \{ k \geq 0:{p^k} < \lambda \}$. Furthermore,
\begin{equation*}
n\left( \lambda \right) =\frac{dN\left( \lambda \right) }{d\lambda }=\left(
1-\frac{1}{\nu}\right) \left[ \delta _{1}\left( \lambda \right) +\frac{\delta
_{p}\left( \lambda \right) }{\nu}+\frac{\delta _{p^{2}}\left( \lambda \right)
}{\nu^{2}}+\cdots \right].
\end{equation*}

\end{proposition}

\begin{proposition}\label{pr3} As $\lambda \downarrow 0$,
\begin{equation*}
N\left( \lambda \right) \asymp\lambda
^{s_{h}/2},~~s_{h}=\frac{2\ln\nu}{\ln(1/p)},
\end{equation*}
or, more precisely,
\begin{equation*}
N\left( \lambda \right) \sim \lambda ^{s_{h}/2}h\left( \frac{\ln \lambda }
{\ln p}\right) .
\end{equation*}
for a positive, periodic function $h(z)={\nu^{-1-\{z\}}}\equiv h(z+1)$. Here, $\{z\}$ is the fractional part of a number $z\in\mathbb{R}$.
\end{proposition}

The latter proposition is a consequence of the following simple calculation. If $[z]$ is the integer part of $z\in R$, then
$$
N\left( \lambda \right) =e^{-k_0(\lambda)\ln\nu}=e^{-\left[ \frac{\ln \lambda}{\ln p}+1
\right]\ln\nu }=e^{- \frac{\ln \lambda}{\ln p}
\ln\nu }e^{(-\left\{ \frac{\ln \lambda}{\ln p}
\right\}-1)\ln\nu }=\lambda ^{s_{h}/2}h\left( \frac{\ln\lambda}{\ln p}
\right) .
$$

We will call the constant $s_{h}=\frac{2\ln \nu}{\ln 1/p}$ the \textit{spectral dimension} of the triple $( X,d_{n}( \cdot ,\cdot ),\Delta_{h}) $.

\textbf{Transition probabilities and the resolvent for $\Delta_h$.}
Let $p\left( t,x,y\right) =P_{x}\{ x\left( t\right) =y\} $ be the transition function of the hierarchical random walk $x\left( t\right) $, i.e.,
$$
\frac{\partial p}{\partial t}=\Delta p, ~~p(0,x,y)=\delta_y(x),
$$
and let
\begin{equation*}
R_{\lambda }\left( x,y\right) =\int\limits_{0}^{\infty }e^{-\lambda
t}p\left( t,x,y\right) dt,~~\lambda >0.
\end{equation*}%
The functions $p$ and $R_\lambda$ define the bounded integral operators
\begin{eqnarray*}
\left( P_{t}f\right) \left( x\right) &=&\sum\limits_{y\in X}p\left(
t,x,y\right) f\left( y\right), \\
\left( R_{\lambda }f\right) \left( x\right) &=&\sum\limits_{y\in
X}R_{\lambda }\left( x,y\right) f\left( y\right)
\end{eqnarray*}%
acting in $l^{\infty }\left( X\right) $ and $l^{2}\left( X\right) $, respectively.

Formula (\ref{ddec}) (where each term on the right is an eigenfunction of $\Delta_h$) and the Fourier method lead to the following statement

\begin{proposition}\label{pr4} The transition kernel $p\left( t,x,y\right) $  has
the form:
\begin{equation*}
p\left( t,x,x\right) =\left( 1-\frac{1}{\nu}\right) \left[ e^{-t}+\frac{e^{-pt}
}{\nu}+\cdots +\frac{e^{-p^kt}}{\nu^{k}}+\cdots \right]~~\text{for each}~x\in X.
\end{equation*}
\begin{equation}\label{2p}
p\left( t,x,y\right) =-\frac{e^{-p^{r-1}t}}{\nu^{r}}+\left( 1-\frac{1}{\nu}
\right) \left( \frac{e^{-p^{r}t}}{\nu^{r}}+\frac{e^{-p^{r+1}t}}{\nu^{r+1}}
+\cdots \right),~~x\neq y.
\end{equation}
Here, $r=d_{h}\left( x,y\right) $ is the minimal rank of the cube $Q^{\left( \cdot
\right) }\left( x\right) $, containing the point $y$ $(see (\ref{dh}))$.

\end{proposition}
Similar formulas for $R_\lambda(x,y)$ can be obtained from (\ref{ddec}) or (easier) from the proposition above (by integration in $t$):

\begin{proposition} \label{pr7a} For any $s_{h}>0,\lambda >0,$
\begin{equation*}
R_{\lambda }\left( x_{0},x\right) =-\frac{1}{(\lambda +p^{r-1})\nu^{r}}+\left( 1-\frac{1}{\nu}\right)\left(\frac{1}{(\lambda +p^{r})\nu^{r}}+ \frac{1}{(\lambda +p^{r+1})\nu^{r+1}}
 +\cdots\right),
\end{equation*}%
when $r=d_{h}(x_{0},x)>0.$ If $x_0=x$, then $(independent~of~x\in X)$,
\begin{equation}\label{rxx}
R_\lambda (x,x)=\left(1-\frac{1}{\nu}\right)\left[\frac{1}{\lambda+1}+\frac{1}{(\lambda+p)\nu}+\cdots+\frac{1}{(\lambda+p^s)\nu ^s}+\cdots\right]~~
\end{equation}

\end{proposition}

\begin{corollary}\label{pr5} $(a)$ If $p\nu >1~(s_h=\frac{2\ln \nu}{\ln 1/p}>2)$, then for each $x\in X$,
\begin{equation*}
R_{0}(x,x)=\int\limits_{0}^{\infty}p(t,x,x)dt=\left( 1-\frac{1}{\nu}\right) \left( 1+\frac{1}{p\nu }+\frac{1}{\left( p\nu \right) ^{2}}+\cdots \right)=\frac{p(\nu-1)}{p \nu -1}<\infty.
\end{equation*}
If $p\nu \leq1$ $(i.e.,~s_h=\frac{2\ln \nu}{\ln (1/p)}\leq2)$, then $\lim_{\lambda\to +0}R_\lambda(x,x)=\infty$. Thus the random walk $x(t)$ with the generator $\Delta_{h}$ is transient for $s_{h}>2$ and recurrent for $s_{h}\leq 2$.

$(b)$ If $s_h>2$ and $\rho(x_0,x)\to\infty$ (see (\ref{dist})), then
\begin{eqnarray*}
R_{0}(x_{0},x) &=&\left( \frac{1}{p^{r}\nu^{r}}-\frac{1}{p^{r-1}\nu^{r}}\right) + \left( \frac{1}{p^{r+1}\nu^{r+1}}-\frac{1}{p^{r}\nu^{r+1}}\right) +\cdots \\&=&\frac{1-p}{(p\nu)^{r-1}(p\nu-1)}\sim\frac{c}{\rho^{s_h-2}(x_{0},x)},~~~c= \frac{p\nu(1-p)}{p\nu-1}.
\end{eqnarray*}
This is one more indication of a similarity between $\Delta_{h}$ and the lattice ${\mathbb{Z}^d}$ Laplacian.
\end{corollary}
Now let's find the asymptotics of $p(t,x,x)$ as $t\rightarrow \infty$. The asymptotics will play an essential role in the spectral theory of the Schr\"{o}dinger operator $H=-\Delta_{h}+V(x)$.

\begin{proposition}\label{pr6} For arbitrary spectral dimension $s_{h}$,
\begin{equation*}
p(t,x,x) \asymp\frac{1}{t^{s_{h}/2}},~~t\to\infty,
\end{equation*}
and there exists a positive periodic function $h_1(z)\equiv h_1(z+1)$ such that
\begin{equation}\label{1p}
p(t,x,x)=\frac{h_1\left(\frac{\ln t}{\ln(1/p)}\right)}{t^{s_h/2}}(1+o(1)) \quad \text{as}~~t\to\infty.
\end{equation}
\end{proposition}

\textbf{Proof.} The index of the maximal term in the series $p(t,x,x)=(1-\frac{1}{\nu})\sum_{s=0}^\infty\frac{e^{-p^st}}{\nu^s}$ has order $s=O(\frac{\ln t}{\ln(1/p)})$ when $t\to\infty$. We put $k=[\frac{\ln t}{\ln(1/p)}]$ and change the order of terms in the series representation of $p$, first taking the sum  over $s\geq k$ and then taking the sum over $s<k$:
\begin{equation}\label{pper}
\begin{split}
p(t,x,x)& =\left(1-\frac{1}{\nu}\right)\left(\frac{e^{-p^{k}t}}{\nu^k}+\frac{e^{-p^{(k+1)}t}}{\nu^{k+1}}+\cdots+\frac{e^{-p^{(k-1)}t}}{\nu^{k-1}}+\cdots \right)\\
 & =\left(1-\frac{1}{\nu}\right)\frac{e^{-p^{k} t}}{\nu^k}\left[1+\frac{e^{p^{k}t(1-p)}}{\nu}+\frac{e^{p^{k}t(1-p^2)}}{\nu^2}+\cdots+\frac{e^{p^{k}t(1-\frac{1}{p})}}{\nu^{-1}}+\frac{e^{p^{k}t(1-\frac{1}{p^2})}}{\nu^{-2}}+\cdots \right].\\
\end{split}
\end{equation}
The relation $\frac{\ln t}{\ln(1/p)}=k+\{\frac{\ln t}{\ln(1/p)}\}$ implies that
\begin{equation*}
p^{k}t=p^{-\{\frac{\ln t}{\ln(1/p)}\}} \quad \text{and} \quad \frac{1}{\nu^k}=e^{-\frac{\ln t}{\ln(1/p)}\ln \nu} \nu^{-\{\frac{\ln t}{\ln(1/p)}\}}=\frac{1}{t^{s_h/2}}\nu^{-\{\frac{\ln t}{\ln(1/p)}\}}.
\end{equation*}
We substitute the latter relations into (\ref{pper}) and note that $\{x\}$ is a periodic function of $x$ with period one.

This and (\ref{pper}) would lead to (\ref{1p}) with zero reminder term if both series in square brackets in (\ref{pper}) had infinitely many terms. Since the second part in the square brackets has only $k$ terms we obtain (\ref{1p}) with an exponentially small reminder.
\qed

The next statement provides the asymptotic expansion of $R_\lambda (x,x)$ as $\lambda \to +0$. We restrict ourselves to the more difficult and important case where $s_h<2$. As in the previous proposition, the main term of the expansion contains a periodic function. We will use an alternative approach to show that.

\begin{proposition}\label{pr6} If $s_{h}<2$, then
\begin{equation*}
R_\lambda(x,x)=\lambda^{-\alpha}u\left(\frac{\ln \lambda}{\ln p}\right)+c_0+O(\lambda), \quad \lambda\to +0, ~~\alpha=1-\frac{\ln\nu}{\ln1/p}=1-\frac{s_h}{2},
\end{equation*}
where $c_0=\frac{p(\nu-1)}{p\nu-1}$ is a constant and $u(z+1)=u(z)$ is a positive periodic function with period one.
\end{proposition}
\textbf{Proof.} From series representation (\ref{rxx}) it follows that
$$
R_{p\lambda}-\frac{1}{p\nu}R_\lambda=\frac{\nu-1}{\nu(p\lambda+1)}.
$$
We put $R_\lambda=c_0+f(\lambda)$. Then
 $$
f(p\lambda)-\frac{1}{p\nu}f(\lambda)=\frac{p(1-\nu)}{\nu(p\lambda+1)}\lambda.
$$
After the substitution $f(\lambda)=\lambda^{-\alpha}g(\lambda)$ we arrive at
\begin{equation}\label{perg}
g(p\lambda)-g(\lambda)=\zeta(\lambda), \quad \zeta(\lambda)=\frac{{p^{2}(1 - \nu )}}{p\lambda+1}{\lambda^{1+\alpha}}.
\end{equation}

The estimate $|\zeta(\lambda)|<C|\lambda^{1+\alpha}|, ~\lambda>0,$ is valid for the function $\zeta$ (this estimate was the goal of the subtraction of the constant $c_0$ from $R_\lambda$ made above). Hence the series $g_{\textrm par}=\sum_0^\infty \zeta (p\lambda),~\lambda>0$, converges, has order $O(\lambda^{1+\alpha})$ as $\lambda\to +0$ and is a partial solution of equation (\ref{perg}). Any solution of the homogeneous equation (\ref{perg}) is a periodic function of $\ln_p\lambda=\frac{\ln \lambda}{\ln p}$ with period one. This completes the proof. \qed

\textbf{Remark.} The statement of the proposition and its proof remain valid if $\lambda\to 0$ in the complex plane, and $|\arg \lambda|\leq3\pi/4$.

We conclude this section by defining two functions, $\theta(t)$ and $\varsigma(z)$, which are the analogues of the corresponding classical 1-D functions:
\begin{eqnarray*}
\theta(t) &=&\int\limits_{0}^{\infty }e^{-\lambda
t}dN(\lambda) =\left( 1-\frac{1}{\nu}\right) \left[ e^{-t}+\frac{
e^{-pt}}{\nu}+\frac{e^{-p^{2}t}}{\nu^{2}}+\cdots\right] ,\\
\varsigma(z) &=&\frac{1}{\Gamma(z)}\int\limits_{0 }^{\infty }t^{z-1}\theta(t)dt=\left( 1-\frac{1}{\nu}\right)\sum_{r=0}^{\infty }\frac{1}{p^{rz}\nu^{r}}=\left( 1-\frac{1}{\nu}\right) \frac{p^{z}\nu}{p^{z}\nu-1}.
\end{eqnarray*}

The formula for $\varsigma \left( z\right) $ is obtained for $\operatorname{Re} z\in (0,\delta)$ with a small enough $\delta>0$ ($p^{\operatorname{Re} z}\nu>1$) and understood in the sense of the analytic continuation for other $z$. The function $\varsigma$ has no complex zeros, but (compare to \cite{acc}) has infinitely many poles at $z=z_{n}=\frac{s_{h}}{2}+\frac{i\pi n}{\ln 1/p}$.

\section{Elements of the general lattice spectral theory}

The functions $p(t,x,y)$ and $R_{\lambda}(x,y)$ play a central role in the analysis of the positive spectrum of the
hierarchical Schr\"{o}dinger operator
\begin{equation}\label{11}
H=\Delta_h+V(x);~ V\geq 0.
\end{equation}%
With only weak assumptions on $V$, the positive spectrum $\lambda_{n}=\lambda_{n}(H)\geq 0$ of $H$ is discrete (possibly, with accumulation at $\lambda =0$). Our goals are to find upper bounds on $N_{0}(V)=\#\{\lambda _n\geq 0\}$ and on the Lieb-Thirring sums $S_{\gamma }(V)=\sum\nolimits_{n}(\lambda_{n})^\gamma ,\gamma>0$. Below, we will provide several estimates on $N_{0}$ and $S_{\gamma}$ which are valid \cite{mv},\cite{arxiv} for general discrete operators and for the operator (\ref{11}) in particular.

Let $X$ be an arbitrary countable set and let $H_{0}$ be a bounded self-adjoint operator on $l^{2}(X)$ given by
\begin{equation*}
H_{0}\psi(x)=\sum\limits_{y:y\neq x}h(x,y)\left(\psi(y)-\psi(x)\right) ,
\end{equation*}

\begin{equation*}
h(x,y)=h(y,x) \geq 0 ~~\textrm{for}~~ x\neq y,~~~\sum\limits_{y:y\neq x}h(x,y) \leq C_{0}<\infty.
\end{equation*}
It is clear that $H_{0}=H^{\ast}_{0},~H_{0}\leq 0,~\left\Vert H_{0}\right\Vert \leq 2C_{0}$.

Let $p(t,x,y)=P_{x}\left(x(t) =y\right)$ be the transition kernel of the continuous time Markov chain $x(t)$ generated by $H_{0}$. Of course,
\begin{equation*}
\frac{\partial p}{\partial t}=H_{0}p,~~p(0,x,y)=\delta_{y}(x).
\end{equation*}

We assume that $x(t)$ is connected which means, since its time is continuous, that $p(t,x,y)>0$ for arbitrary $x,y\in X$ and $t>0$.

The bounds for the eigenvalues of $H_{0}$ depend essentially on whether the process $x(t)$ is transient or recurrent. If $\int\nolimits_{0}^{\infty}p(t,x,x)dt<\infty$ for every $x\in X$, then $x(t)$ is transient, i.e., $P$-a.s., $x(t) \rightarrow \infty$ as $t\rightarrow \infty$. If $\int\nolimits_{0}^{\infty}p(t,x,x)dt=\infty$ for every $x\in X$, then $x(t)$ visits each state $x\in X$ infinitely many times $P$-a.s. and the process is called recurrent. It is a well-known fact that, if the chain is connected, the convergence or divergence of $\int\nolimits_{0}^{\infty }p(t,x,y)dt$ is independent of $x,y$.

\begin{theorem}\label{t31h}
$(General~CLR~estimate~for~discrete~operators)$. If $\int\nolimits_{0}^{\infty }p(t,x,x)dt<\infty$, then for any $a,\sigma>0$ and some $c_1(\sigma)$,
\begin{equation*}
N_{0}(V) \leq \#\{ x\in X:V(x)>a\}+c_{1}(\sigma)\sum\limits_{x:V(x) \leq a}V(x) \int\limits_{\frac{\sigma }{V(x)}}^{\infty}p(t,x,x)dt
\end{equation*}
\end{theorem}

\begin{theorem}\label{t32h}
$(LT~estimate)$. If $\int\nolimits_{0}^{\infty}p(t,x,x)dt<\infty$ then
\begin{equation*}
S_{\gamma }(V) \leq \frac{1}{c(\sigma)}\sum\limits_{x\in X}V^{1+\gamma }(x)\int\limits_{\frac{\sigma }{V(x)}}^{\infty}p(t,x,x)dt.
\end{equation*}
\end{theorem}

\begin{theorem}\label{t33h}
If $\int\nolimits_{1}^{\infty}t^{-\gamma}p(t,x,x)dt<\infty$ for some $\gamma>0$, then
\begin{equation*}
S_{\gamma }(V) \leq \frac{2\gamma \Gamma(\gamma) }{c(\sigma)}\sum\limits_{x\in X}V(x)\int\limits_{\frac{\sigma }{V(x)}}^{\infty}t^{-\gamma}p(t,x,x)dt.
\end{equation*}
$(Note~that~here,~the~process~x(t)~may~not~be~transient)$.
\end{theorem}

The following two results are valid in both transient and recurrent cases. These results are based on
the method of partial annihilation, proposed in \cite{mv},\cite{arxiv}. In the
discrete situation it is equivalent to the rank-one perturbation technique.

Consider, for a fixed $x_{0}\in X$, the process $x(t)$ with the condition of annihilation at $x_{0}$. The corresponding transition
probability $p_{1}( t,x,y)$ is given by
\begin{equation}\label{h1}
\frac{\partial p_{1}}{\partial t}=H_{0}p_{1},~~x,y\neq x_{0}, \quad p_1(t,x_{0},y)\equiv 0;~~~\quad  ~~p_1(0,x,y)=\delta_{y}(x).
\end{equation}
As easy to see, $\int\nolimits_{0}^{\infty }p_{1}(t,x,x)dt<\infty$.

\begin{theorem}\label{t34h}
$(CLR~estimate,~the~general~case)$. For any $a,\sigma>0$ and some $c_1(\sigma)$,
\begin{equation*}
N_{0}(V)\leq 1+\#\{x:V(x)>a\}+c_{1}(\sigma)\sum\limits_{x:V(x)\leq a}V(x)\int\limits_{\frac{\sigma}{V(x)}}^{\infty}p_{1}(t,x,x)dt.
\end{equation*}
\end{theorem}

\begin{theorem}\label{t35h}
$(LT~estimates,~the~general~case)$. The following two estimates hold for each $\sigma\geq 0$ and some $c(\sigma)>0$:
\begin{equation} \label{lit9}
S_\gamma(V)\leq\Lambda^\gamma+\frac{1}{c(\sigma)}\sum_X V^{1+\gamma}(x)\int_{\frac{\sigma}{V(x)}}^\infty p_1(t,x,x)dt,
\end{equation}
\begin{equation} \label{lithi9}
S_\gamma(V)\leq\Lambda^\gamma+\frac{2\gamma \Gamma(\gamma)}{c(\sigma)}\sum_X V(x)\int_{\frac{\sigma}{V(x)}}^\infty t^{-\gamma}p_1(t,x,x)dt.
\end{equation}
Here $\Lambda$ is the largest eigenvalue of $H$.
\end{theorem}

Note that Theorem \ref{t34h} not only covers the recurrent case, but also provides a better result than Theorems \ref{t31h}, \ref{t32h} in the transient case when the operator $H=H^{\alpha}$ depends on a parameter $\alpha$ which approaches a threshold $ \alpha=\alpha_0$, where the process becomes recurrent. In Theorems \ref{t31h}, \ref{t32h}, the integrals in $t$ blow up when $\alpha$ approaches $\alpha_0$  whereas they remain bounded in Theorem \ref{t34h}. A similar remark is valid for Theorem \ref{t35h} where the threshold depends on the values of $\alpha$ and $\gamma$.

In the case where $\sigma=0$, \cite{arxiv} contains a more detailed description of the results obtained in Theorems \ref{t31h}-\ref{t35h}.

\section{The perturbation of the hierarchical Laplacian.}

Theorems \ref{t31h}-\ref{t33h} and Proposition \ref{pr6}, when applied to the operator (\ref{11}), lead to the same bounds on $N_0(V)$ and $S_\gamma(V)$ as in the case of the standard Schr\"{o}dinger operator in ${\mathbb{R}^d}$ with the dimension $d$ replaced by the spectral dimension $s_h$. An essential difference is that, while $d$ must be an integer, the spectral dimension $s_h$ can be an arbitrary positive number. The corresponding bounds hold if $s>2$, where $s=s_h$ in the estimate on $N_0(V)$ and $s=\gamma+\frac{s_h}{2}$ in the estimates on $S_\gamma(V)$. The right-hand sides in these estimates blow up when $s \downarrow 2$ (the integrals in $t$ diverge when $s=2$). For example, Theorem \ref{t31h} with $\sigma =0$ and Proposition \ref{pr6} imply a usual estimate:
\[
N_0(V)\leq \#\{x\in X:V(x)>a\}+\frac{C(A)}{s_h-2} \sum\limits_{x:V(x) \leq a}V^{s_h/2}(x), \quad 2<s_h<A.
\]

The case $s\leq 2$ is covered by Theorems \ref{t34h}, \ref{t35h}. In fact, these theorems are valid for any $s>0$ and the estimates proven there are (locally) uniform in $s$. Hence they provide a better result in the transient case $s>2$ than do Theorems \ref{t31h}-\ref{t33h} when $s \downarrow 2$, see \cite{arxiv}.

In order to apply Theorems \ref{t34h}, \ref{t35h}, one needs to know an estimate on $p_1$ as $t\to\infty$ and both the annihilation point $x_0$ and $x$ are arbitrary. If $\sigma=0$, then only the integral $\int_0^\infty p_1dt$  is needed, not $p_1$ itself. The corresponding results can be found in \cite{arxiv} (we concentrated on $N_0(V)$ in  \cite{arxiv}, but $S_\gamma(V)$ can be studied similarly). Theorem \ref{t34h} with $\sigma=0$ implies \cite{arxiv} the following Bargmann type result:
\begin{equation}\label{1z1}
N_0(V)\leq 1+\#\{x:~V(x)\geq 1\}+C_1(s_h)\sum_{x:V(x)<1}V(x)\rho(x_0,x)^{2-s_h}, \quad s_h< 2,
\end{equation}
with $C_1(s_h)\to \infty$ as $s_h\to 2$. A more accurate estimate of $\int_0^\infty p_1dt$ leads  \cite{arxiv} to estimates on $N_0(V)$ for all $s_h$ and with a uniformly bounded constant:
\begin{theorem} If $\varepsilon< s_h<\varepsilon^{-1},~s_h\neq 2$, then
\begin{equation}\label{1z1}
N_0(V)\leq 1+\#\{x:~V(x)\geq 1\}+C_2(\varepsilon)\sum_{x:V(x)<1}V(x)\frac{[1+\rho(x_0,x)]^{2-s_h}-1}{(\frac{1}{\sqrt{p}})^{2-s_h}-1},
\end{equation}
If $s_h= 2,$ then
$$
N_0(V)\leq 1+\#\{x:~V(x)\geq 1\}+C_2\sum_{x:V(x)<1}V(x)\frac{\ln [1+\rho(x_0,x)]}{\ln\frac{1}{\sqrt p}}.
$$
\end{theorem}

In this section, we will obtain an estimate for $p_1$ as $t\to\infty$, which allows one to use Theorems \ref{t34h}, \ref{t35h} with arbitrary $\sigma>0$. We will restrict ourselves to the case where $s_h<2$ and provide an estimate only on $N_0(V)$. The following refined Bargmann type estimate is an immediate consequence of Theorem \ref{t34h} and Proposition \ref{prlast} which will be proven below:
\begin{theorem} If $s_h< 2$, then
$$
N_0(V)\leq 1+\#\{x:~V(x)\geq 1\}+C_1(s_h)\sum_{x:V(x)<1}V^{2-\frac{s_h}{2}}(x)[1+\rho^2(x_0,x)]^{2-s_h} .
$$
\end{theorem}
We will conclude this section with a proof of the estimate on $p_1$ as $t\to\infty$. This estimate is needed to justify the refined Bargmann estimate stated above and to prove similar estimates for $S_\gamma$.
\begin{proposition}\label{prlast}
The following estimate is valid
\[
p_1(t,x,x)\leq C\frac{(\rho^2+1)^{2\alpha}}{t^{1+\alpha}}, \quad t\geq 1, \quad  \rho=\rho(x_0,x), \quad  \alpha= 1-\frac{s_h}{2}.
\]
\end{proposition}
\textbf{Remark.} We expect that, in the case of fractal lattices similar to the Sierpincki lattice, the same estimate will be valid for a random walk with annihilation at a point.

\textbf{Proof.} Consider the function
\begin{equation} \label{rpp}
R^{(1)}_\lambda(x,y)=\int_0^\infty e^{-\lambda t}p_1(t,x,y)dt.
\end{equation}
It is well defined when $\operatorname{Re}\lambda>0$ and understood in the sense of analytic continuation for complex $\lambda \in C_+=\{\lambda\in \mathbb{C}:|\textrm{arg}\lambda|<3\pi/4\}$. From (\ref{h1}) it follows that $R^{(1)}_\lambda$ satisfies
$$
(\Delta_h-\lambda)R^{(1)}_\lambda(x,y)=-\delta_y(x), ~~x,y\neq x_{0}, \quad R^{(1)}_\lambda(x_{0},y)= 0.
$$
Hence $R^{(1)}_\lambda(x,y)=R_\lambda(x,y)+cR_\lambda(x,x_0)$, which together with the second relation in the formula above implies that
\[
R^{(1)}_\lambda(x,y)=R_\lambda(x,y)-\frac{R_\lambda(x_0,y)}{R_\lambda(x_0,x_0)}R_\lambda(x,x_0).
\]

We put here $y=x$ and $R_\lambda(x_0,x)=R_\lambda(x_0,x_0)+\widetilde{R}_\lambda(x_0,x)$ where (see Proposition \ref{pr7a})
\begin{equation} \label{rwa}
\widetilde{R}_\lambda(x_0,x)=-\frac{1}{(\lambda +p^{r-1})\nu^{r}}-(1-\frac{1}{\nu})\sum_{s=0}^{r-1}\frac{1}{(\lambda +p^{s})\nu^{s}},  \quad r=d_h(x_0,x).
\end{equation}
Taking also into account that $R_\lambda(x,x_0)=R_\lambda(x_0,x)$ and $R_\lambda(x,x)$ does not depend on $x$, we obtain that
\begin{equation} \label{rrr}
R^{(1)}_\lambda(x,x)=-2\widetilde{R}_\lambda(x_0,x)-\frac{\widetilde{R}_\lambda^2(x_0,x)}{R_\lambda(x_0,x_0)}.
\end{equation}
We note that (\ref{rwa}) immediately implies the following two estimates:
\[
|\widetilde{R}_\lambda(x_0,x)|\leq \frac{c}{(p\nu)^r}, \quad  |\widetilde{R}_\lambda(x_0,x)-\widetilde{R}_0(x_0,x)|\leq \frac{c|\lambda|}{(p\nu)^r} \quad \textrm {for all}~~\lambda\in C_+,~r\geq 0,
\]
which together with (\ref{rrr}) and the Remark after Proposition \ref{pr6} lead to
\begin{equation} \label{rez}
R^{(1)}_\lambda(x,x)=a(r)+g(\lambda,r), \quad a(r)=-2\widetilde{R}_\lambda(x_0,x),~~|g|\leq \frac{2c|\lambda|}{(p\nu)^r}+\frac{c_1|\lambda|^{\alpha}}{(p\nu)^{2r}}.
\end{equation}
The last estimate is valid for all $\lambda\in C_+$ with $|\lambda|<1$ and all $r\geq 0$.

Applying the inverse Laplace transform to (\ref{rpp}) we obtain
\[
p_{1}(t,x,x)=\frac{1}{2\pi}\int_{b-i\infty}^{b+i\infty} e^{\lambda t}R^{(1)}_\lambda(x,x)d\lambda,~~b\gg1.
\]
Since $R^{(1)}_\lambda$ is analytic in $\lambda\in C_+$, and $|R^{(1)}_\lambda|\leq \frac{1}{|\textrm{Im}\lambda|}$ (the resolvent does not exceed the inverse distance from the spectrum), the last integral can be rewritten as
\[
p_1(t,x,x)=\frac{1}{2\pi}\int_{\Gamma} e^{\lambda t}R^{(1)}_\lambda(x,x)d\lambda,
\]
where $\Gamma=\partial C_+$ with the direction on $\Gamma$ such that Im$\lambda$ increases along $\Gamma$. We now use (\ref{rez}), the decay of $R_\lambda^{(1)}$ on $\Gamma$ at infinity, and the fact that $\int_{\Gamma} e^{\lambda t}d\lambda=0,~t>0.$ This leads to
\[
p_1(t,x,x)\leq\frac{1}{2\pi}\int_{\Gamma} |e^{\lambda t}|\left(\frac{2c|\lambda|}{(p\nu)^r}+\frac{c_1|\lambda|^{\alpha}}{(p\nu)^{2r}}\right)|d\lambda|=\frac{a_1}{t^2(p\nu)^r}+
\frac{a_2}{t^{1+\alpha}(p\nu)^{2r}}.
\]

It remains to recall that $\alpha=1-\frac{\ln\nu}{\ln1/p}$ (see Proposition \ref{pr6}). Thus $p\nu=p^\alpha$, and $\frac{1}{(p\nu)^r}=\frac{1}{p^{\alpha r}}=(\rho^2+1)^\alpha$.
\qed


\begin{thebibliography}{10}

\bibitem{Dy} F. Dyson, Existance of a phase transition in a one-dimensional Ising ferromagnetic, Comm. Math. Phys., V. 12,(1969), no. 2, 91-107.

\bibitem{md} S. Molchanov, Hierarchical random matricies and operators. Application to Anderson model, Proc. of Sixth Lukacs Symp., VSP (1996), pp179-194.

\bibitem{mv} S. Molchanov, B. Vainberg, On general Cwikel-Lieb-Rozenblum and Lieb-Thirring inequalities, in "Around the research of Vladimir Maz'ya", III, Editor A. Laptev, Int. Math. Ser. (N.Y.) 13, Springer, 2010, pp 201-246.

\bibitem{arxiv} S. Molchanov, B. Vainberg, On negative eigenvalues of low-dimensional Schr\"odinger operators, arXiv:1105.0937

\bibitem{kig} J.  Kigami, : Analysis on Fractals. Cambridge Univ. Press, Cambridge, 2001.

\bibitem{fuk} M. Fukushima, T. Shima "On a spectral analysis for Sierpinski gasket",
Potential Anal. 1(1992), \#1, pp 1-35.

\bibitem{mal} L. Malozemov, A. Teplyaev "Pure point spectrum of the Laplacian on
fractal graphs", F. Funct. Anal. 129 (1995), \#2, 396-405.

\bibitem{acc}
 E. Akkermans, G. Dunne, A. Teplyaev, Thermodynamics of photons on fractals,  Phys.Rev.Lett.105:230407,(2010).




\bibitem{bov} A Bovier,The density of states in the Anderson model at weak disorder: a renormalization
group analysis of the hierarchical model, J. Statist. Phys. 59 (1990), no. 3-4,
745�779.

\bibitem{kri} E. Kritchevski, Spectral Localization in the Hierarchical Anderson Model,
Proceedings of the American Mathematical Society
135, No 05, (2005) 1431-1441



\end{thebibliography}
\end{document}